\documentclass[12pt]{article}
\usepackage{amsmath,amssymb}
\usepackage{graphicx,color}
\numberwithin{equation}{section}
\usepackage{cite}
\usepackage{bm}
\usepackage{dcolumn}
\newcommand{\be}{\begin{equation}}
\newcommand{\bea}{\begin{eqnarray}}
\newcommand{\eea}{\end{eqnarray}}
\newcommand{\ba}{\begin{array}}
\newcommand{\ea}{\end{array}}
\newcommand{\ee}{\end{equation}}

\expandafter\ifx\csname mathbbm\endcsname\relax

\else

\fi \textheight 22cm \textwidth 15cm \topmargin 1mm \oddsidemargin
5mm \evensidemargin 5mm

\begin{document}
\begin{titlepage}
\hfill \vbox{
    \halign{#\hfil         \cr
         \cr
                      } 
      }  
\vspace*{20mm}
\begin{center}
{\Large {\bf The Constrained State of Minimum Energy and the Effective Equation of Motion}\\
}

\vspace*{15mm} \vspace*{1mm} {Amin Akhavan}

\vspace*{.4cm}

{\it  School of Particles and Accelerators, Institute for Research in Fundamental Sciences (IPM)\\
P.O. Box 19395-5531, Tehran, Iran\\$email:amin_- akhavan@ipm.ir$ }

\vspace*{2cm}

\end{center}

\begin{abstract}
We define the state of minimum energy while the expectation values of the field operators and their time derivatives in a determined moment in such a state are constrained. As an axiom, we consider such a state as the background of the quantum field theory. As an example, we consider the scalar field with $\frac{\lambda}{4!}\Phi^4$ interaction. To the third order of perturbation, we obtain the equation of motion of the dynamic expectation value of the scalar field in the defined state.
\end{abstract}

\vspace{2cm}

\end{titlepage}

\section{Introduction}
 One of the interesting interpretations of effective actions has been introduced by K.Symanzik. He defined the effective action as the minimum of the Hamiltonian expectation value, while the expectation values of field operators have been constrained\cite{zymanski}. In this interpretation, the vacuum state will be defined as a spontaneous state in which the effective action is stationary and at a minimum. An important consequence of this method is that the effective potential curve in the vacuum point must be convex. The symmetry-breaking method was obtained from this viewpoint.\\

 In the case of the renormalization conditions with $m^2<0$, the effective action has two minimum points with a negative-definite second derivative between them. But if the expectation value of the field is a value between these minimum points (in a non-zero source), the effective action has a concave curve, which is in contradiction to the minimum energy interpretation. To solve this contradiction, one has to consider the effective potential as a constant value between the two minimum points, which satisfies the effective potential to be convex\cite{weinberg}.\\

In continuing, one can define an axiom in which all the ground states are in minimum energy while the expectation values of the field operators in these states have to be equal to the background external fields.\\

In the Symanzik method, the Lagrange multipliers mechanism is used such that the non-dynamic multipliers are related to the sources which are inside the generating functions. In this method, the field operators' expectation values, which satisfy the equations with the sources, must also satisfy the equations without the sources. Therefore, the expectation values will not be dynamic.\\
\be
<\Omega_{J}|\Phi(t,\vec{x})|\Omega_{J}>=<\Omega_{J}|\Phi_{J}(t,\vec{x})|\Omega_{J}>=\phi_{J}(\vec{x}).
\ee
If the expectation value of a field operator were to be dynamic, the Lagrange multipliers could not be related completely to the sources in the generator function $W(J)$, and the effective action could not be proportional to the hamiltonian expectation value.\\

Pay attention to the macroscopic external fields that are measured classically. If such external fields are not dynamic, we can consider them as the expectation values of the field operators in the vacuum state\cite{faddeev}:\\
\be
\phi(\vec{x})=<\Omega|\Phi(t,\vec{x})|\Omega>.
\ee
In this consideration, the non-dynamic external fields satisfy the equations which are derived from the variation of the effective action. But they are trivial solutions that have no application for the measured macroscopic fields. Because the measured dynamic external fields have to be obtained from the excited states except for the vacuum.\\

The variation of the effective actions determines only the sources ($J=-\frac{\delta \Gamma}{\delta \phi}$). For example, the variation of the effective actions determines that the sources of the vacuum state are zero, and in the background field method, the vacuum expectation value of the fields should be the constant mean fields that satisfy the classical equation\cite{faddeev}.\\ 

Here, we must carefully note that the effective action is only a generating function, not a function for deriving the equation of motion of the measured external fields. In this paper, Similar to Symanzik's method, we want to define the background state in the presence of the source, except that the time evolution operator will be the source-free Hamiltonian. Our goal is to find dynamic background fields and their equations of motion. For this purpose, the important questions are:\\ 1. In which quantum state does the measurement of the dynamic external fields occur?\\ 2. What is the amount of the expectation values of field operators in this state?\\3. what are the equations of motion of these external fields?\\

In this manuscript, in the second section, we define the axiom of minimum energy for the dynamic external fields. In the third section and its subsection, we obtain the equation of motion for the dynamic external scalar field. In the last section, we discuss introducing the spontaneous excited state and explaining the future works of this research.
\section{The constrained state of minimum energy}
We define a spontaneous state $|\psi>$ in which the expectation values of the field operators are considered as the external fields:
\be
\phi(t,\vec{x})=<\psi|\Phi(t,\vec{x})|\psi>.
\ee
For a dynamic external field, we can not consider $|\psi>$ as a hamiltonian eigenstate and we need to use a method to derive it. We slightly expand the axiom of the minimum energy which has been presented by Symanzik. We determine the external field and also its time derivative in a special fixed time $t_{0}$ as the constraints inside the axiom:
\be
\phi(t_{0},\vec{x})=\varphi_{0}(\vec{x})~~~~~\dot{\phi}(t_{0},\vec{x})=\dot{\varphi}_{0}(\vec{x}).
\ee
Considering the above constraints and also $<\psi|\psi>=1$, and by using the Lagrange multipliers method, we minimize the expectation value of the Hamiltonian:
\bea
\delta <\psi|H|\psi>-\int d^3xJ_{0}(\vec{x}) \delta<\psi|\Phi(t_{0},\vec{x})|\psi>-\int d^3xK_{0}(\vec{x})\delta<\psi|\dot{\Phi}(t_{0},\vec{x})|\psi>\nonumber\\
-E\delta<\psi|\psi>=0~~~~~~~~~~~~~~~~~~~~~~~~~~~~
\eea
Therefore:
\be
\int d^3x\bigg(\mathcal{H}[\Phi(t_{0},\vec{x}),\dot{\Phi}(t_{0},\vec{x})]-J_{0}(\vec{x})\Phi(t_{0},\vec{x})-K_{0}(\vec{x})\dot{\Phi}(t_{0},\vec{x})\bigg)|\psi>=E|\psi>.
\ee
In equation (2.4), we can see that $|\psi>$ is the minimum eigenstate of a new operator which is defined at time $t_{0}$. We call this operator, $H_{S(Source)}$.  \\
For the models where the hamiltonian has a quadratic functional of $\dot{\Phi}(t_{0},\vec{x})$, we can write the operator defined above as follows:
\be
H_{S}=\int d^3x\mathcal{H}_{S}=\int d^3x \frac{1}{2}\pi^2(\vec{x})+V[\Phi(t_{0},\vec{x})]-J_{0}\Phi(t_{0},\vec{x})
\ee
while
\be
\pi(\vec{x})=\dot{\Phi}(t_{0},\vec{x})-K_{0}(\vec{x})
\ee
and
\be
[\Phi(t_{0},\vec{x}),~\pi(\vec{x'})]=i\delta^3(\vec{x}-\vec{x'}).
\ee
According to the appearance of $H_{S}$, we conclude that:
\be
<\psi|\pi|\psi>=0
\ee
and therefore:
\be
K_{0}(\vec{x})=\dot{\varphi}_{0}(\vec{x}).
\ee
In the following, we have to consider the right-hand side of equation (2.1) as an n-point function of operators at time $t_{0}$ in the state $|\psi>$.  Although we have defined $|\psi>$ as an eigenstate of the sourced Hamiltonian, we base the time evolution of the operators on the sourceless Hamiltonian, thus from equation (2.1), we have:\\
\be
\phi(t,\vec{x})=<\psi|\Phi(t,\vec{x})|\psi>=<\psi|e^{i(t-t_{0})H}\Phi(t_{0},\vec{x})e^{-i(t-t_{0})H}|\psi>.
\ee

To obtain this type of n-point function at $t_{0}$, through the Feynman diagrams, we have to consider the state $|\psi>$ as like as below:
\be
\lim_{S \to \infty (1+i\epsilon)} e^{iSH_{S}}|0>=\lim_{S \to \infty (1+i\epsilon)} e^{iSE}|\psi><\psi|0>,
\ee
where the state $|0>$ is the vacuum eigenstate of the $H_{S}$ with $V[\Phi(t_{0},\vec{x})]=0$ . Also, we define a Heisenberg picture along a hypothetical parameter $s$ like this:
\be
\Phi_{S}(s,\vec{x})=e^{isH_{S}}\Phi(t_{0},\vec{x})e^{-isH_{S}}\nonumber\\
\ee
\be
\pi_{S}(s,\vec{x})=e^{isH_{S}}\pi(\vec{x})e^{-isH_{S}}=\frac{d\Phi_{S}(s,\vec{x})}{ds}.
\ee
In this case, all the n-point functions can be written like this:
\be
<\psi|A[\Phi(t_{0},\vec{x})]|\psi>=<\psi|e^{isH_{S}}A[\Phi(t_{0},\vec{x})]e^{-isH_{S}}|\psi>\nonumber\\
\ee
\be
=<\psi|A[\Phi_{S}(s,\vec{x})]|\psi>=\frac{<0|A[\Phi_{I}(s,\vec{x})]e^{-i\int dsd^3x\mathcal{H}_{S,I}}|0>}{<0|e^{-i\int dsd^3x\mathcal{H}_{S,I}}|0>}
\ee
in which $\pi_{I}$ and $\Phi_{I}$ are the operators in the interaction picture along the\\ parameter $s$.\\
As we know:
\be
<\psi|\Phi_{S}(s,\vec{x})]|\psi>=<\psi|\Phi(t_{0},\vec{x})]|\psi>=\varphi_{0}(\vec{x})
\ee
For the above equation to be true, we will obtain the source $J_{0}(\vec{x})$,
\be
J_{0}(\vec{x})=-\frac{\delta \Gamma[\varphi_{0}(\vec{x})]}{\delta \varphi_{0}(\vec{x})}.
\ee
$ \Gamma(\varphi_{0})$ is the effective action for the external fields whose coordinates are only spatial. We can use the source with spatial coordinates, to find $\mathcal{H}_{S}(\vec{x})$ in the equation (2.5).

In the following, we have to consider the equation of motion of $\phi(t,\vec{x})$. Since the equation of motion for the operator $\Phi(t,\vec{x})$ is :
\be
\ddot{\Phi}(t,\vec{x})-\nabla^2\Phi(t,\vec{x})+\frac{\delta V[\Phi(t,\vec{x})]}{\delta \Phi(t,\vec{x})}=0,
\ee
therefore the expectation value of this equation at time $t_{0}$, is written as follows:
\be
<\psi|\ddot{\Phi}(t_{0},\vec{x})|\psi>-\nabla^2\varphi_{0}(\vec{x})+<\psi|\frac{\delta V[\Phi(t_{0},\vec{x})]}{\delta\Phi(t_{0},\vec{x})}|\psi>=0.
\ee
By using the equation (2.13 ), the n-point function $<\psi|\frac{\delta V[\Phi]}{\delta\Phi}|\psi>$ will be obtained. We can assume another time $t_{1}$ with these constraints:
\be
<\psi_{1}|\Phi(t_{1},\vec{x})|\psi_{1}>=\varphi_{1}(\vec{x})~~~~~~~~~<\psi_{1}|\dot{\Phi}(t_{1},\vec{x})|\psi_{1}>=\dot{\varphi}_{1}(\vec{x}).            
\ee
We can reproduce (2.17) with the state $|\psi_{1}>$. Using equation (2.10), we can choose $\varphi_{1}(\vec{x})=\phi(t_{1},\vec{x})$ and $\dot{\varphi}_{1}(\vec{x})=\dot{\phi}(t_{1},\vec{x})$ and this choice leads to:
\be
|\psi_{1}>=|\psi>
\ee
Therefore we have equation (2.17) for all times, with a constant state (Heisenberg picture):
\be
\ddot{\phi}(t,\vec{x})-\nabla^2\phi(t,\vec{x})+<\psi|\frac{\delta V[\Phi(t,\vec{x})]}{\delta\Phi(t,\vec{x})}|\psi>=0.
\ee
\section{Scalar field example}
In this section, we are going to obtain the equation of motion for a scalar field with an interaction term $\frac{\lambda}{4!}\Phi^4$.\\
By considering the Lagrangian density,
\bea
\mathcal{L}=\frac{1}{2}\partial_{\mu}\Phi \partial^{\mu}\Phi-\frac{1}{2}m^2\Phi^2-\frac{\lambda}{4!}\Phi^4\nonumber\\
+\frac{1}{2}\delta_{z}\partial_{\mu}\Phi \partial^{\mu}\Phi-\frac{1}{2}\delta_{m}\Phi^2-\frac{\delta_{\lambda}}{4!}\Phi^4
\eea
and by using equation(2.4), we have:
\bea
\mathcal{H}_{S}=\frac{1}{2}\dot{\Phi}^2(t_{0},\vec{x})+\frac{1}{2}(\vec{\nabla}\Phi(t_{0},\vec{x}))^2+\frac{1}{2}m^2\Phi^2(t_{0},\vec{x})+\frac{\lambda}{4!}\Phi^4(t_{0},\vec{x})\nonumber\\
-\frac{1}{2}\delta_{z}\partial_{\mu}\Phi(t_{0},\vec{x})\partial^{\mu}\Phi(t_{0},\vec{x})+\frac{1}{2}\delta_{m}\Phi^2(t_{0},\vec{x})+\frac{\delta_{\lambda}}{4!}\Phi^4(t_{0},\vec{x})\nonumber\\
-J_{0}(\vec{x})\Phi(t_{0},\vec{x})-K_{0}(\vec{x})\dot{\Phi}(t_{0},\vec{x})-\delta J_{0}(\vec{x})\Phi(t_{0},\vec{x})-\delta K_{0}(\vec{x})\dot{\Phi}(t_{0},\vec{x}).
\eea
We will rewrite the field operators as below,
\be
\Phi(t_{0},\vec{x})=\varphi_{0}(\vec{x})+\chi(\vec{x})\nonumber\\
\ee
\be
\dot{\Phi}(t_{0},\vec{x})=\dot{\varphi}_{0}(\vec{x})+\pi(\vec{x})
\ee
such that the constraints will be written like this:
\be
<\psi|\chi(\vec{x})|\psi>=0 ~~and ~~<\psi|\pi(\vec{x})|\psi>=0.
\ee
By using the second constraint and also considering
\bea
K_{0}(\vec{x})=\dot{\varphi}_{0}(\vec{x})\nonumber\\
\delta K_{0}(\vec{x})=-\delta_{z}\dot{\varphi}_{0}(\vec{x}) \nonumber\\
J_{0}(\vec{x})=-\nabla^2\varphi_{0}(\vec{x})+m^2\varphi_{0}(\vec{x}),
\eea
we will have:
\bea
\mathcal{H}_{S}=\frac{1}{2}\pi^2(\vec{x})+\frac{1}{2}(\nabla\chi(\vec{x}))^2+\frac{1}{2}m^2\chi^2(\vec{x})+\frac{\lambda}{4!}(\varphi_{0}(\vec{x})+\chi(\vec{x}))^4\nonumber\\
-\frac{1}{2}\delta_{z}\pi^2(\vec{x})+\frac{1}{2}\delta_{z}(\nabla\chi(\vec{x}))^2+\delta_{z}\nabla\chi(\vec{x}).\nabla\varphi_{0}(\vec{x})\nonumber\\
+\frac{1}{2}\delta_{m}(\varphi_(\vec{x})+\chi(\vec{x}))^2+\frac{\delta_{\lambda}}{4!}(\varphi_{0}(\vec{x})+\chi(\vec{x}))^4-\delta J_{0}(\vec{x})\chi(\vec{x})+const,
\eea

such that in the interaction picture along the parameter $s$, we can write the interaction term as below:
\bea
\mathcal{H}_{S,I}(s,\vec{x})=\frac{\lambda}{4!}(\varphi_{0}(\vec{x})+\chi_{I}(s,\vec{x}))^4
-\frac{1}{2}\delta_{z}(\frac{d\chi_{I}(s,\vec{x})}{ds})^2 \nonumber\\+\frac{1}{2}\delta_{z}(\nabla\chi_{I}(s,\vec{x}))^2+\delta_{z}\nabla\chi_{I}(s,\vec{x}).\nabla\varphi_{0}(\vec{x})\nonumber\\
+\frac{1}{2}\delta_{m}(\varphi_{0}(\vec{x})+\chi_{I}(s,\vec{x}))^2+\frac{\delta_{\lambda}}{4!}(\varphi_{0}(\vec{x})+\chi_{I}(s,\vec{x}))^4\nonumber\\-\delta J_{0}(\vec{x})\chi_{I}(s,\vec{x})+const.
\eea
To find~ $\delta J_{0}(\vec{x})$, we need to use the first constraint:
\be
<\psi|\chi(\vec{x})|\psi>=<\psi|\chi_{S}(s,\vec{x})|\psi>=0.
\ee
(Where $\chi_{S}(s,\vec{x})$ is the Heisenberg picture along the parameter $s$.)
To do so, we have to use the equation (2.13) to the first order of the interaction coefficient $\lambda$.
Therefore we consider the odd power terms of $\chi$ in $\mathcal{H}_{S}$ to write the expectation value to the first order:
\bea
<\psi|\chi_{S}(s,\vec{x})|\psi>^{(\lambda)}=\int ds'd^3x'\frac{\lambda+\delta_{\lambda}^{(\lambda)}}{3!}\varphi^3_{0}(\vec{x'})<0|\chi_{I}(s,\vec{x})\chi_{I}(s',\vec{x'})|0>\nonumber\\+3\int ds'd^3x'\frac{\lambda+\delta_{\lambda}^{(\lambda)}}{3!}\varphi_{0}(\vec{x'})<0|\chi_{I}(s',\vec{x'})\chi_{I}(s',\vec{x'})|0><0|\chi_{I}(s,\vec{x})\chi_{I}(s',\vec{x'})|0>\nonumber\\-\delta_{z}^{(\lambda)}\int ds'd^3x'\nabla^2\varphi_{0}(\vec{x'})<0|\chi_{I}(s,\vec{x})\chi_{I}(s',\vec{x'})|0>+\delta_{m}^{(\lambda)}\int ds'd^3x'\varphi_{0}(\vec{x'})<0|\chi_{I}(s,\vec{x})\chi_{I}(s',\vec{x'})|0>\nonumber\\
-\int ds'd^3x'\delta J_{0}^{(\lambda)}(\vec{x'})<0|\chi_{I}(s,\vec{x})\chi_{I}(s',\vec{x'})|0>=0~~~~~~~~~~~~
\eea
and we will obtain:
\bea
\delta J_{0}^{(\lambda)}(\vec{x'})=\frac{\lambda+\delta_{\lambda}^{(\lambda)}}{3!}\varphi^3_{0}(\vec{x'})+\frac{\lambda+\delta_{\lambda}^{(\lambda)}}{2}\varphi_{0}(\vec{x'})G(0)
-\delta_{z}^{(\lambda)}\nabla^2\varphi_{0}(\vec{x'})+\delta_{m}^{(\lambda)}\varphi_{0}(\vec{x'}).~~~~~~~~
\eea	
In continue, we write the equation (2.17) for this example:
\bea
\ddot{\phi}(t_{0},\vec{x})-\nabla^2\varphi_{0}(\vec{x})+m^2\varphi_{0}(\vec{x})+\frac{\lambda+\delta_{\lambda}}{3!}<\psi|(\varphi_{0}(\vec{x})+\chi(\vec{x}))^3|\psi>\nonumber\\
+\delta_{z}\ddot{\phi}(t_{0},\vec{x})-\delta_{z}\nabla^2\varphi_{0}(\vec{x})+\delta_{m}\varphi_{0}(\vec{x})=0.
\eea
Since
\be
<\psi|(\varphi_{0}(\vec{x})+\chi(\vec{x}))^3|\psi>=\varphi^3_{0}(\vec{x})+3\varphi_{0}(\vec{x})<\psi|\chi^2(\vec{x})|\psi>
+<\psi|\chi^3(\vec{x})|\psi>,\nonumber\\
\ee
and its zeroth order is $\varphi^3_{0}(\vec{x})+3\varphi_{0}(\vec{x})G(0)$, therefore the equation (3.11) to the first order is:
\bea
\ddot{\phi}(t_{0},\vec{x})-\nabla^2\varphi_{0}(\vec{x})+m^2\varphi_{0}(\vec{x})+\frac{\lambda+\delta_{\lambda}^{(\lambda)}}{3!}(\varphi^3_{0}(\vec{x})+3\varphi_{0}(\vec{x})G(0))\nonumber\\
+\delta_{z}^{(\lambda)}\ddot{\phi}(t_{0},\vec{x})-\delta_{z}^{(\lambda)}\nabla^2\varphi_{0}(\vec{x})+\delta_{m}^{(\lambda)}\varphi_{0}(\vec{x})=0.
\eea
To keep the recent equation finite, we have to choose the counter term coefficients in the equation (3.12) like this:
\be
\delta_{\lambda}^{(\lambda)}=\delta_{z}^{(\lambda)}=0~~ and ~~\delta_{m}^{(\lambda)}=-\frac{\lambda}{2}G(0),
\ee
and therefore, equation (3.10) changes to this:
\be
\delta J_{0}^{(\lambda)}(\vec{x})=\frac{\lambda}{3!}\varphi^3_{0}(\vec{x}).
\ee
\subsection{The higher order of perturbations}
To the second order of pertubation, we can write equation (3.11) like this:
\bea
\ddot{\phi}(t_{0},\vec{x})-\nabla^2\varphi_{0}(\vec{x})+m^2\varphi_{0}(\vec{x})+\frac{\lambda}{3!}\varphi^3_{0}+\frac{\delta_{\lambda}^{(\lambda^2)}}{3!}(\varphi^3_{0}(\vec{x})+3\varphi_{0}(\vec{x})G(0))\nonumber\\
+\frac{\lambda}{3!}(3\varphi_{0}(\vec{x})<\psi|\chi^2(\vec{x})|\psi>^{(\lambda)}+<\psi|\chi^3(\vec{x})|\psi>^{(\lambda)})\nonumber\\
+(-\delta_{z}^{(\lambda^2)}m^2+\delta_{m}^{(\lambda^2)})\varphi_{0}(\vec{x})=0.
\eea
In continue, for simplicity, we consider the background field as a constant in the spatial coordinate. From (3.7), to the first order of perturbation, we have:
\bea
\mathcal{H}^{(\lambda)}_{S,I}(s,\vec{x})=\frac{\lambda}{4!}\chi^4_{I}(s,\vec{x})+\frac{\lambda}{3!}\varphi_{0}\chi^3_{I}(s,\vec{x})+\frac{\lambda}{4}\varphi^2_{0}\chi^2_{I}(s,\vec{x}) 
 \nonumber\\
+\frac{1}{2}\delta^{(\lambda)}_{m}\chi^2_{I}(s,\vec{x})+\delta^{(\lambda)}_{m}\varphi_{0}\chi_{I}(s,\vec{x})\nonumber\\+const.
\eea
Using equation (2.13), we have:
\bea
<\psi|\chi^2(\vec{x})|\psi>^{(\lambda)}=\frac{i\lambda\varphi^2_{0}}{2}\mathcal{A}_{2}=\frac{i\lambda\varphi^2_{0}}{2}\int\frac{d^4p}{(2\pi)^4}\frac{-1}{(p^2-m^2+i\epsilon)^2}.
\eea
\bea
<\psi|\chi^3(\vec{x})|\psi>^{(\lambda)}=i\lambda\varphi_{0}\mathcal{A}_{3}=~~~~~~~~~~~~~~~~~~~~~~~~~~~~~~~~~~~~~~~~~~~~\nonumber\\i\lambda\varphi_{0}\int\frac{d^4p}{(2\pi)^4}\frac{d^4k}{(2\pi)^4}\frac{i}{p^2-m^2+i\epsilon}\frac{i}{k^2-m^2+i\epsilon}\frac{i}{(p+k)^2-m^2+i\epsilon}.
\eea
As you can see, the second-order expressions are divergent and proportional to $\varphi_{0}$ and  $\varphi^3_{0}$. Therefore the equation of motion in the second order will be trivialized as in the first order. 
Using equation (3.8), we can find the second-order source:
\bea
\delta J_{0}^{(\lambda^2)}=\delta_{m}^{(\lambda^2)}\varphi_{0}+\delta_{\lambda}^{(\lambda^2)}G(0)\frac{\varphi_{0}}{2}+\frac{i\lambda^2\varphi_{0}}{3!}\mathcal{A}_{3}.
\eea 
To find non-trivial expressions, we have to examine the third order. The third order of the equation of motion is:
\bea
e.q.^{(\lambda^3)}=\frac{\lambda}{3!}<\psi|(\varphi_{0}+\chi(\vec{x}))^3|\psi>^{(\lambda^2)}+\frac{\delta_{\lambda}^{(\lambda^2)}}{3!}<\psi|(\varphi_{0}(\vec{x})+\chi(\vec{x}))^3|\psi>^{(\lambda)}\nonumber\\+\frac{\delta_{\lambda}^{(\lambda^3)}}{3!}(\varphi_{0}^3+3G(0)\varphi_{0})-\delta_{z}^{(\lambda^2)}\frac{\lambda\varphi_{0}^3}{3!}-\delta_{z}^{(\lambda^3)}m^2\varphi_{0}+\delta_{m}^{(\lambda^3)}\varphi_{0}~~~~~~~~~~~~\nonumber\\=\frac{\lambda}{3!}\bigg(3\varphi_{0}<\psi|\chi^2(\vec{x})|\psi>^{(\lambda^2)}+<\psi|\chi^3(\vec{x})|\psi>^{(\lambda^2)}\bigg)~~~~~~~~~~~~~~~~~~~~~~~~~~~~~~~~\nonumber\\+\bigg(\frac{\delta_{\lambda}^{(\lambda^3)}}{3!}+\frac{i\lambda\delta_{\lambda}^{(\lambda^2)}}{4}\mathcal{A}_{2}-\frac{\lambda\delta_{z}^{(\lambda^2)}}{3!}\bigg)\varphi_{0}^3~~~~~~~~~~~~~~~~~~~~~~~~~~~~~~~~~~~~~~~~~\nonumber\\
+\bigg(-m^2\delta_{z}^{(\lambda^3)}+\delta_{m}^{(\lambda^3)}+\frac{\delta_{\lambda}^{(\lambda^3)}}{2}G(0)+\frac{i\lambda\delta_{\lambda}^{(\lambda^2)}}{3!}\mathcal{A}_{3}\bigg)\varphi_{0}.~~~~~~~~~~~~
\eea
To find the expectation values in the above equation we need to obtain $\mathcal{H}_{S, I}^{(\lambda^2)}$. By putting $\delta J_{0}^{(\lambda^2)}$ in the equation (3.7), we will have:
\bea
\mathcal{H}_{S, I}^{(\lambda^2)}=\frac{1}{2}\chi_{I}(s,\vec{x})(\delta_{z}^{(\lambda^2)}\partial^2+\delta_{m}^{(\lambda^2)})\chi_{I}(s,\vec{x})+\frac{\delta_{\lambda}^{(\lambda^2)}}{4!}\chi_{I}^4(s,\vec{x})+\frac{\delta_{\lambda}^{(\lambda^2)}}{3!}\varphi_{0}\chi_{I}^3(s,\vec{x})\nonumber\\+\frac{\delta_{\lambda}^{(\lambda^2)}}{4}\varphi_{0}^2\chi_{I}^2(s,\vec{x})+\frac{\delta_{\lambda}^{(\lambda^2)}}{3!}\varphi_{0}^3\chi_{I}(s,\vec{x})-(\frac{\delta_{\lambda}^{(\lambda^2)}}{2}G(0)+\frac{i\lambda^2\mathcal{A}_{3}}{3!})\varphi_{0}\chi_{I}(s,\vec{x}).~~
\eea
For $<\psi|\chi^3(\vec{x})|\psi>^{(\lambda^2)}$, we have:
\bea
<\psi|\chi^3(\vec{x})|\psi>^{(\lambda^2)}=~~~~~~~~~~~~~~~~~~~~~~~~~~~~~~~~~~~~~~~~~~~~~~~~~~~~~~~~~~~~~~~~~~~~~~\nonumber\\<0|\chi_{I}^3(s,\vec{x})\bigg(i\int d^4y\mathcal{H}_{S,I}^{(\lambda^2)}(y)+\frac{i^2}{2}\int d^4yd^4z\mathcal{H}_{S,I}^{(\lambda)}(y)\mathcal{H}_{S,I}^{(\lambda)}(z)\bigg)|0>_{connected}.
\eea
As you can see, the sentences containing $\chi_{I}$ with odd power, in $\mathcal{H}_{S,I}^{(\lambda^2)}$ and $\mathcal{H}_{S,I}^{(\lambda)}\mathcal{H}_{S,I}^{(\lambda)}$, are proportional to $\varphi^3_{0}$ and $\varphi_{0}$.
But for $\varphi_{0}<\psi|\chi^2(\vec{x})|\psi>^{(\lambda^2)}$ we can find a term in which is not proportional to $\varphi^3_{0}$ or $\varphi_{0}$. In even power of $\chi_{I}$ in $\mathcal{H}_{S,I}^{(\lambda)}\mathcal{H}_{S,I}^{(\lambda)}$, we have, $\frac{\lambda^2}{16}\varphi_{0}^4\chi^2_{I}(y)\chi^2_{I}(z)$. Thus, in $e.q.^{(\lambda^3)}$ we have this convergent term:
\bea
-\frac{\lambda^3}{64}\varphi_{0}^5\int d^4yd^4z<0|\chi_{I}^2(s,\vec{x})\chi^2_{I}(y)\chi^2_{I}(z)|0>_{connected}=\nonumber\\\frac{i\lambda^3\varphi_{0}^5}{8}\int\frac{d^4p}{(2\pi)^4}\frac{1}{(p^2-m^2+i\epsilon)^3}=\frac{\lambda^3}{(16\pi)^2m^2}\varphi_{0}^5.
\eea
As the renormalization condition, we eliminate the divergent  $\varphi_{0}$-term and the  $\varphi^3_{0}$-term . From equation (2.20) we have:
\be
\ddot{\phi}(t)+m^2\phi(t)+\frac{\lambda}{3!}\phi^3(t)+\frac{\lambda^3}{(16\pi)^2m^2}\phi^5(t)+...=0.
\ee
 
\section{Discussion}
In a quantum mechanical system, a ground state is the eigenstate of Hamiltonian with the minimum eigenvalue. In normal modes, such systems are in the ground states unless provoked, then enter the excited states. The excited states have small lifetimes and the systems move back to the ground states as soon as possible. But in quantum field theory, we need excited states that are stable for long times, states that describe background fields and their interactions in addition to the existing particles. We may call them spontaneous excited states. These states have minimum energy while they are not the ground states of the hamiltonians. This spontaneous excitation is related to the sources which have been added to the hamiltonian. But contrary to the De Witt mechanism\cite{Dewitt}, these sources have no role in the equations of motions of the field operators. As a result, there is no source in the equation of motion of the external fields.\\

In this paper, the equation of motion was obtained to the third order of perturbation. Here we chose simple boundary conditions in which the background field was constant in the spatial coordinate.
In the next research, we will compute the equation of motion for any boundary conditions. For this purpose, we need to explore different renormalization conditions.\\

Here we defined the constraints at the determined moment. In the future, we will check if any constraints could be defined at the moment and in a determined spacial spot. In fact, in minimizing the Hamiltonian density at a given space-time point, we will investigate if the expectation values of the fields can be constrained. We need to find out if there are local excitations in the field theory.

\end{document}